\documentclass[aps,prd,preprint,superscriptaddress]{revtex4}
\usepackage{graphicx}
\usepackage{bm}
\usepackage{amsmath}
\usepackage{url}

\def\lsim{\mathrel{\rlap{\lower4pt\hbox{\hskip1pt$\sim$}}
    \raise1pt\hbox{$<$}}}         
\def\gsim{\mathrel{\rlap{\lower4pt\hbox{\hskip1pt$\sim$}}
    \raise1pt\hbox{$>$}}}         

\begin{document}


\title{Neutrino Luminosity and Matter-Induced Modification of Collective Neutrino Flavor Oscillations in Supernovae}




\author{John F. Cherry}
\affiliation{Department of Physics, University of California, San Diego, La Jolla, California 92093, USA}
\affiliation{Neutrino Engineering Institute, New Mexico Consortium, Los Alamos, New Mexico 87545, USA}
\author{Meng-Ru Wu}
\affiliation{School of Physics and Astronomy, University of Minnesota, Minneapolis, MN 55455, USA}
\author{J. Carlson}
\affiliation{Theoretical Division, Los Alamos National Laboratory, Los Alamos, New Mexico 87545, USA}
\affiliation{Neutrino Engineering Institute, New Mexico Consortium, Los Alamos, New Mexico 87545, USA}
\author{Huaiyu Duan}
\affiliation{Department of Physics and Astronomy, University of New Mexico, Albuquerque, New Mexico 87131, USA}
\affiliation{Neutrino Engineering Institute, New Mexico Consortium, Los Alamos, New Mexico 87545, USA}
\author{George M. Fuller}
\affiliation{Department of Physics, University of California, San Diego, La Jolla, California 92093, USA}
\affiliation{Neutrino Engineering Institute, New Mexico Consortium, Los Alamos, New Mexico 87545, USA}
\author{Yong-Zhong Qian}
\affiliation{School of Physics and Astronomy, University of Minnesota, Minneapolis, MN 55455, USA}

\date{\today}

\begin{abstract}
We show that the bump in the electron number density profile 
at the base of the hydrogen envelope in O-Ne-Mg core-collapse 
supernovae causes an interesting interplay 
between neutrino-electron and neutrino-neutrino 
forward scattering effects in the flavor evolution of low-energy $\nu_e$ 
in the neutronization burst.
The bump allows a significant fraction of the low-energy $\nu_e$ 
to survive by rendering their flavor evolution nonadiabatic.
Increasing the luminosity of the neutronization burst
shifts the bump-affected $\nu_e$ to lower energy with reduced
survival probability.  Similarly, lowering the luminosity shifts the bump-affected neutrinos
to higher energies.  While these low energy neutrinos lie near the edge of detectability, the
population of bump-affected neutrinos has direct influence on the spectral swap formation 
in the neutrino signal at higher energies.  
\end{abstract}

\maketitle

\section{Introduction}
Stars of $\sim 8$--$10\,M_\odot$ ($M_\odot$ being the mass of the sun)
develop O-Ne-Mg cores at the end of their evolution. Capture of electrons
by the Ne and Mg isotopes initiates the gravitational collapse of the core,
which eventually produces a supernova and leaves behind a neutron star~\cite{Nomoto84,Nomoto87}. 
These O-Ne-Mg core-collapse
supernovae are the only case for which the neutrino-driven mechanism
has been demonstrated to work by different groups~\cite{Mayle88,Kitaura06}. The success
of this mechanism is largely due to the steep fall-off of the matter density 
above the core. As shown by Refs.~\cite{Duan08,Dasgupta:2008qy,Cherry:2010lr,Cherry:2011bg}, this special density structure also provides a venue where neutrino flavor transformation occurs
under the influence of both neutrino-electron and
neutrino-neutrino forward scatterings~\cite{Fuller87,Notzold:1988fv,Pantaleone92,Fuller:1992eu,Qian93,Samuel:1993sf,Qian95,Kostelecky:1995rz,Samuel:1996rm,Pastor02,Pastor:2002zl,Sawyer:2005yg,Fogli07,Duan08,Dasgupta:2008qy,Duan:2010fr,Cherry:2010lr,Cherry:2011bg}.  In particular,
the neutronization burst, which consists predominantly of $\nu_e$ and
signifies the breakthrough of the neutrino sphere by the supernova shock,
experiences interesting flavor evolution including collective oscillations
for the neutrino flavor mixing parameters found by experiments~\cite{Nakamura:2010lr}.

In this paper we explore another special feature of the matter structure in O-Ne-Mg core-collapse supernovae in connection with flavor 
evolution of the neutronization neutrino burst. The hydrogen envelope
has an electron fraction of $Y_e\approx 0.85$. In contrast, the material
below the envelope has $Y_e\approx 0.5$, reflecting weak interaction-induced neutronization during pre-supernova evolution. As
the matter density $\rho$ is a continuous function of radius, this produces
a bump in the electron number density $n_e=\rho Y_eN_A$ 
($N_A$ being Avogadro's number) at the base of the hydrogen envelope, shown explicitly in Ref.~\cite{Cherry:2011bg}.  As noted in Refs.~\cite{Duan08,Dasgupta:2008qy,Cherry:2010lr,Cherry:2011bg}, for the normal neutrino mass hierarchy,
this bump renders flavor evolution of
the low-energy $\nu_e$ in the neutronization burst
nonadiabatic, giving rise to substantial survival
probabilities for these $\nu_e$. Here we show that this bump facilitates
an interesting interplay between neutrino-electron and
neutrino-neutrino forward scattering in the flavor evolution
of the low-energy $\nu_e$, and we show how this influences the collective oscillations
 of neutrinos at higher energies.  

\section{Neutronization Burst Neutrinos}

We assume a pure $\nu_e$ burst emitted from the neutrino sphere at
$R_\nu=60$~km with a total luminosity $L_{\nu}=10^{52}-10^{54}\,{\rm erg}\,{\rm s}^{-1}$
and a normalized spectrum
\begin{equation}
f_{\nu}(E)=\frac{1}{F_2(\eta_{\nu})T_{\nu}^3}
\frac{E^2}{\exp(E/T_{\nu}-\eta_{\nu})+1},
\end{equation}
where we take $\eta_{\nu}=3$ and $T_{\nu}=2.75$~MeV [corresponding to an average
$\nu_e$ energy $\langle E_{\nu}\rangle=F_3(\eta_{\nu})T_{\nu}/F_2(\eta_{\nu})=11$~MeV
at emission]. Here
\begin{equation}
F_n(\eta)=\int_0^\infty\frac{x^n}{\exp(x-\eta)+1}dx.
\end{equation}
In the single-angle approximation, the neutrino-neutrino forward scattering potential can be written in terms of an \emph{effective} total neutrino number density at $r>R_\nu$,
\begin{equation}
n_\nu(r)=\frac{L_{\nu}}{4\pi R_\nu^2\langle E_{\nu}\rangle}
\left[1-\sqrt{1-(R_\nu/r)^2}\right]^2\approx
\frac{L_{\nu}R_\nu^2}{16\pi\langle E_{\nu}\rangle r^4},
\end{equation}
where the approximate equality holds for $r\gg R_\nu$.

For the purposes of this study we have chosen the following neutrino mixing parameters: neutrino mass squared differences $\Delta m^{2}_{\odot} = 7.6\times10^{-5}\, \rm eV^{2}$ and $\Delta m^{2}_{\rm atm} = 2.4\times10^{-3}\, \rm eV^{2}$; vacuum mixing angles $\theta_{12} = 0.59$, $\theta_{23} = \pi/4$, $\theta_{13} = 0.1$; and CP-violating phase $\delta = 0$.  We have also chosen to use the single-angle approximation for these calculations, where neutrinos emitted at different angles relative to the surface of the neutrino sphere are assumed to have the same flavor evolution history as neutrinos from a single, representative emission trajectory.

Here we will concentrate on the normal neutrino mass hierarchy, because previous work has shown that mixing at the $\Delta m^{2}_{\rm atm}$ scale with this heirarchy produces interesting collective neutrino flavor oscillations in O-Ne-Mg core-collapse supernovae~\cite{Duan08,Dasgupta:2008qy,Cherry:2010lr,Cherry:2011bg}.  For the atmospheric neutrino mass doublet with an inverted hierarchy, neutrinos in the neutronization burst do not experience any flavor transformation. 

\begin{figure*}[h]
\centering
\includegraphics[scale=.65]{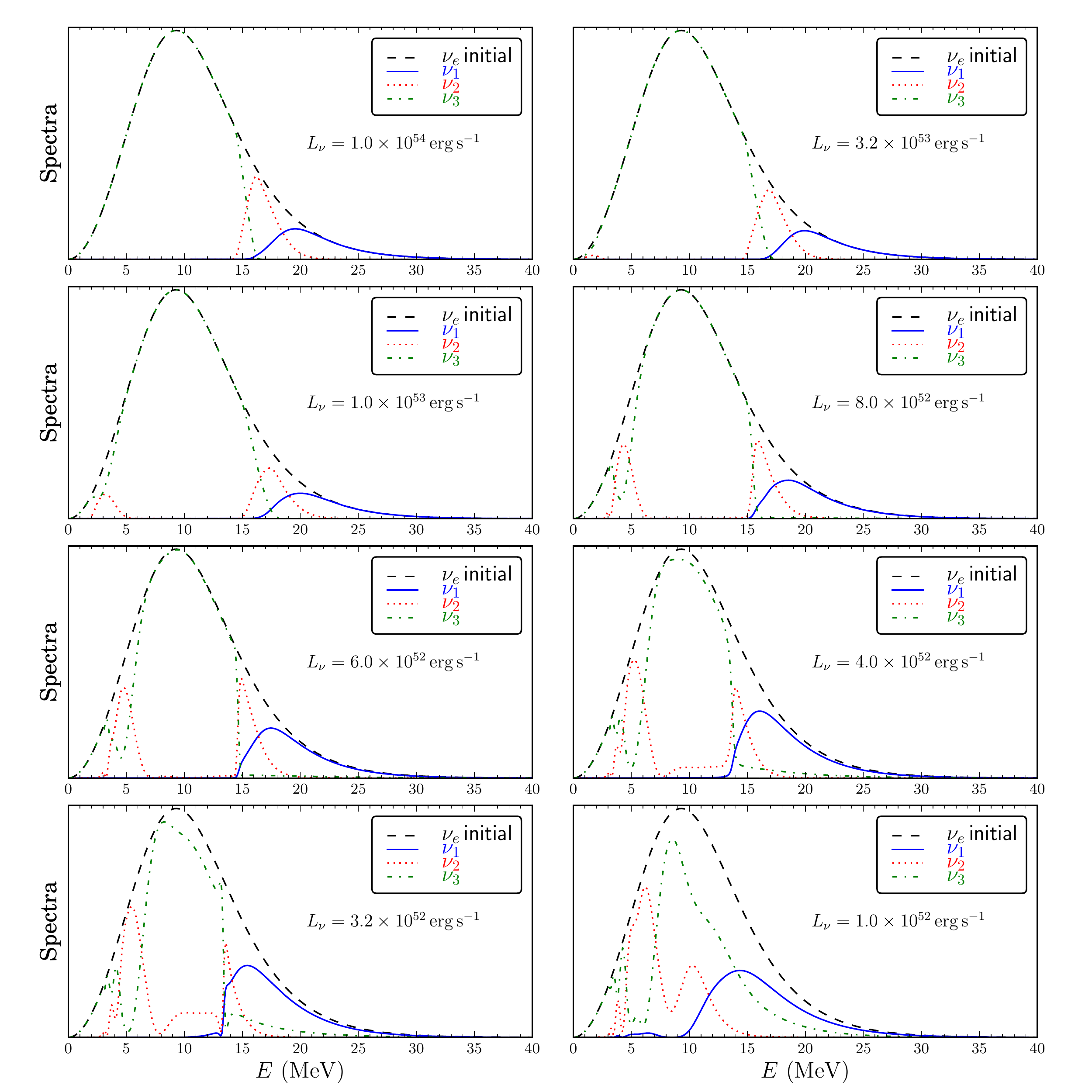}
\caption{The final neutrino mass state emission energy spectra for calculations of the flavor transformation in the neutronization neutrino burst of an O-Ne-Mg core-collapse supernova.  Each panel shows the results for a different possible burst luminosity, ranging from $L_{\nu} = 10^{54} - 10^{52}\,\rm erg\,\rm s^{-1}$, with identical Fermi-Dirac energy distributions.}
\label{fig:AllLum}
\end{figure*}

\begin{figure*}[h]
\centering
\includegraphics[scale=.65]{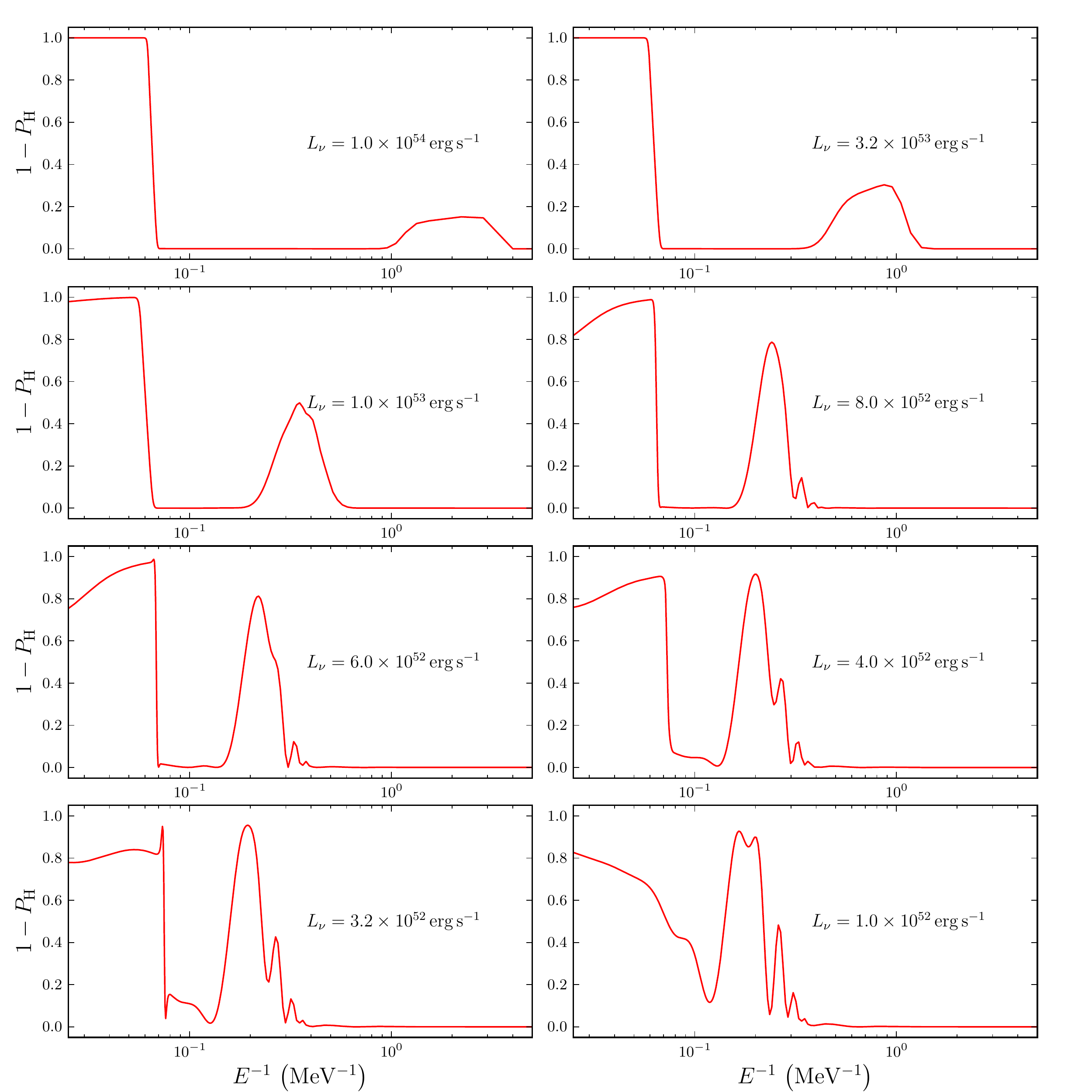}
\caption{The probability for electron neutrinos in the neutronization neutrino burst of an O-Ne-Mg core-collapse supernova to hop out of the (initial) heavy mass eigenstate, $1-P_{\rm H}$, plotted as a function of inverse neutrino energy.  Each panel shows the results for a different possible burst luminosity, ranging from $L_{\nu} = 10^{54} - 10^{52}\,\rm erg\,\rm s^{-1}$, with identical Fermi-Dirac energy distributions.}
\label{fig:AllPH}
\end{figure*}

The results from single-angle simulations shown in Figures~\ref{fig:AllLum} and~\ref{fig:AllPH}
demonstrate that the $n_e$ profile with the bump gives rise to a very
clear flavor transformation signature.  Depending on the luminosity, a population of $\nu_e$ below $8\, \rm MeV$ have large probabilities to transform between neutrino mass states (have large hopping probabilities).  For example, the medium range luminosity case, $L_{\nu} = 8.0\times 10^{52}\,\rm erg\,\rm s^{-1}$, exhibits a peak hopping probability for the bump-affected neutrinos of $\sim 80\%$ at $E_{\nu} = 4.5\,\rm MeV$, shown in Figure~\ref{fig:AllPH}.  In the extreme case of $L_{\nu} = 10^{54}\,\rm erg\,\rm s^{-1}$, the hopping probability is $\sim 15\%$ for $E_{\nu} = 0.5\,\rm MeV$, although this is hard to see in Figure~\ref{fig:AllLum} (but see Figure~\ref{fig:AllPH}). 

\section{Analysis of Flavor Evolution of the $\nu_e\,\rm {'s}$}
As discussed in Refs.~\cite{Duan08,Dasgupta:2008qy,Cherry:2010lr,Cherry:2011bg} the flavor
evolution of the $\nu_e$ flux in this case is governed by 
$\delta m_{\rm atm}^2$ and $\theta_{13}$.   Although the numerical calculations we have conducted employ full $3\times 3$ flavor mixing, in the following analysis,
we focus on the 2-flavor mixing for these $\nu_e$
with $\delta m^2=\delta m_{\rm atm}^2$
and $\theta_{\rm v}=\theta_{13}$. Further, we adopt the single-angle 
approximation, as this has been shown to be surprisingly effective 
in providing qualitative understanding
of the results from multiangle simulations.

Using the notation introduced
in Ref.~\cite{Duan06c}, we can represent a
$\nu_e$ of energy $E$ by a neutrino flavor isospin (NFIS) 
$\mathbf{s}_\omega$ with $\omega=\delta m^2/2E$.
The evolution of $\mathbf{s}_\omega$ is governed by
\begin{equation}
\frac{d}{dr}\mathbf{s}_\omega=\mathbf{s}_\omega\times\left[\omega\mathbf{H}_{\rm v}
+\mathbf{H}_e-\mu(r)\int_0^\infty\mathbf{s}_{\omega'}f_{\nu}(E_{\omega'})dE_{\omega'}\right],
\label{FullEOM}
\end{equation}
where $\mathbf{H}_{\rm v}=\cos2\theta_{\rm v}\mathbf{\hat e}_z^{\rm f}-
\sin2\theta_{\rm v}\mathbf{\hat e}_x^{\rm f}$, 
$\mathbf{H}_e=-\sqrt{2}G_Fn_e(r)\mathbf{\hat e}_z^{\rm f}$, 
$\mu(r)=2\sqrt{2}G_Fn_\nu(r)$,
and $E_{\omega'}=\delta m^2/2\omega'$.  Here $\mathbf{\hat e}_x^{\rm f}$ and $\mathbf{\hat e}_z^{\rm f}$ are the unit vectors in the x and z directions, respectively, of the neutrino flavor space.  For convenience, we define
\begin{equation}
g(\omega)\equiv\frac{\delta m^2}{2\omega^2} f_\nu(E_\omega)
\end{equation}
and
\begin{equation}
\mathbf{S}\equiv\int_0^\infty\mathbf{s}_{\omega}f_{\nu}(E_\omega)dE_\omega=\int_0^\infty\mathbf{s}_{\omega}g(\omega)d\omega.
\label{Sdef}
\end{equation}
It follows that
\begin{equation}
\frac{d}{dr}\mathbf{S}=\int_0^\infty\omega g(\omega)\mathbf{s}_\omega d\omega\times
\mathbf{H}_{\rm v}+\mathbf{S}\times\mathbf{H}_e.
\label{dSdef}
\end{equation}
As $g(\omega)$ is concentrated in a finite range of $\omega$, to zeroth order we
approximate $g(\omega)\approx \delta(\omega-\langle\omega\rangle)$, where
$\langle\omega\rangle=\int_0^\infty\omega g(\omega)d\omega$ is calculated
from the actual $g(\omega)$ in Eq. (5).
Then the zeroth-order mean field $\mathbf{S}^{(0)}$ can be obtained from
\begin{equation}
\frac{d}{dr}\mathbf{S}^{(0)}=\mathbf{S}^{(0)}\times
\left[\langle\omega\rangle\mathbf{H}_{\rm v}+\mathbf{H}_e\right]
\equiv\mathbf{S}^{(0)}\times\mathbf{H}_{\rm MSW}.
\label{S0}
\end{equation}
The evolution of $\mathbf{S}^{(0)}$ is the same as that of a $\nu_e$ with
$E_{\rm MSW}=\delta m^2/2\langle\omega\rangle=8.53$~MeV 
undergoing the usual MSW effect.
With this, we can approximately solve the evolution of $\mathbf{s}_\omega$ by employing
\begin{equation}
\frac{d}{dr}\mathbf{s}_\omega\approx\mathbf{s}_\omega\times\left[\omega\mathbf{H}_{\rm v}
+\mathbf{H}_e-\mu(r)\mathbf{S}^{(0)}\right].
\label{NBEEOM}
\end{equation}

As the heavy mass eigenstate essentially coincides with $\nu_e$ at
high densities, but the light mass eigenstate is predominantly $\nu_e$
at low densities, the survival probability of an initial $\nu_e$ is
approximately $1-P_{\rm H}$, where $P_{\rm H}$ is the probability
for remaining in the heavy mass eigenstate.  

\subsection{Dependence on $L_\nu$} 

We can go further by using the zeroth-order mean field $\mathbf{S}^{(0)}$ to
understand how the flavor evolution of the low-energy $\nu_e$ depends on $L_\nu$.  As discussed below, Equations~\ref{S0} and~\ref{NBEEOM}
imply that neutrinos with $\omega \gg \langle\omega\rangle$ will experience an MSW resonance before the resonance of $\mathbf S^{(0)}$.  These higher frequency neutrinos may pass through multiple resonances created by the matter potential bump.  

Based on the MSW effect, 
$\mathbf{S}^{(0)}$ corresponding to $E_{\rm MSW}=8.53$~MeV
goes through the resonance after the low-energy $\nu_e$.
Assuming adiabatic evolution of $\mathbf{S}^{(0)}$
before the resonance, we can take
\begin{equation}
\mathbf{S}^{(0)}\approx -\frac{\mathbf{H}_{\rm MSW}}{2|\mathbf{H}_{\rm MSW}|}
\approx -\frac{1}{2}(\cos2\theta_m\mathbf{\hat e}_z^f-\sin2\theta_m\mathbf{\hat e}_x^f),
\end{equation}
where
\begin{eqnarray}
\cos2\theta_m&=&\frac{\langle\omega\rangle\cos2\theta_{\rm v}-\sqrt{2}G_Fn_e}
{\sqrt{(\langle\omega\rangle\cos2\theta_{\rm v}-\sqrt{2}G_Fn_e)^2+(\langle\omega\rangle\sin2\theta_{\rm v})^2}},
\label{cos2thm} \\
\sin2\theta_m&=&\frac{\langle\omega\rangle\sin2\theta_{\rm v}}
{\sqrt{(\langle\omega\rangle\cos2\theta_{\rm v}-\sqrt{2}G_Fn_e)^2+(\langle\omega\rangle\sin2\theta_{\rm v})^2}}.
\end{eqnarray}
The evolution of $\mathbf{s}_\omega$ before the resonance of $\mathbf{S}^{(0)}$ is
then governed by
\begin{eqnarray}
\frac{d}{dr}\mathbf{s}_\omega&\approx&\mathbf{s}_\omega\times\left[
(\omega\cos2\theta_{\rm v}-\sqrt{2}G_Fn_e+\frac{\mu}{2}\cos2\theta_m)\mathbf{\hat e}_z^f
-(\omega\sin2\theta_{\rm v}+\frac{\mu}{2}\sin2\theta_m)\mathbf{\hat e}_x^f\right] 
\label{Homega} \\
&\equiv&\mathbf{s}_\omega\times\mathbf{H}_\omega.
\end{eqnarray}
The above equation shows that $\mathbf{s}_\omega$ goes through the resonance when
\begin{equation}
\omega\cos2\theta_{\rm v}=\sqrt{2}G_Fn_e-\frac{\mu}{2}\cos2\theta_m
\equiv |\mathrm{H}_e|+B.
\label{RESCOND}
\end{equation}
Note that $\cos2\theta_m<0$ before the resonance of $\mathbf{S}^{(0)}$
[see Eq. (11)] and therefore $B>0$. Consequently, the energy of those $\nu_e$ 
($E_\omega=\delta m^2/2\omega$) that go through
the resonance at the bump in the $n_e$ profile decreases as $L_\nu$,
and hence $B$, increases.  This trend can be seen in every frame in Figures~\ref{fig:AllLum} and~\ref{fig:AllPH}.  As the neutrino luminosity is increased, the peak energy of the population of low energy neutrinos that hop out of the heavy neutrino mass eigenstate as a result of the bump decreases.

We can also qualitatively understand why increasing $L_\nu$ produces a decreasing survival probability
of the bump-affected $\nu_e$.
The Landau-Zener probability for hopping from the heavy to the light
mass eigenstate after the resonance is
\begin{equation}
P_{\rm hop}=\exp\left[-\frac{\pi}{4}
\frac{\delta m^2\sin^22\theta_{\rm v}}{E\cos2\theta_{\rm v}}{\cal H}_{\rm res}\right],
\end{equation}
where
\begin{equation}
{\cal H}_{\rm res}\equiv\left|\frac{d\ln (|\mathrm{H}_e|+B)}{dr}\right|_{\rm res}^{-1}
\end{equation}
is the scale height of the total flavor-evolution potential at the resonance position. Crudely we have $1-P_{\rm H}\sim P_{\rm hop}$.
As increasing $L_\nu$ shifts the resonance energy
window to lower $E_\nu$ at the bump in the $n_e$ profile, $P_{\rm hop}$
decreases because flavor evolution through the resonance tends to
be more adiabatic for lower-energy neutrinos [see Eq. (16)]. In addition, 
as $B$ decreases much more slowly than $|\mathrm{H}_e|$ with radius,
${\cal H}_{\rm res}$ becomes larger when the contribution from $B$
increases with $L_\nu$. This also reduces $P_{\rm hop}$ [see Eq. (16)].

\begin{figure}[htp]
\centering
\includegraphics[scale=0.7]{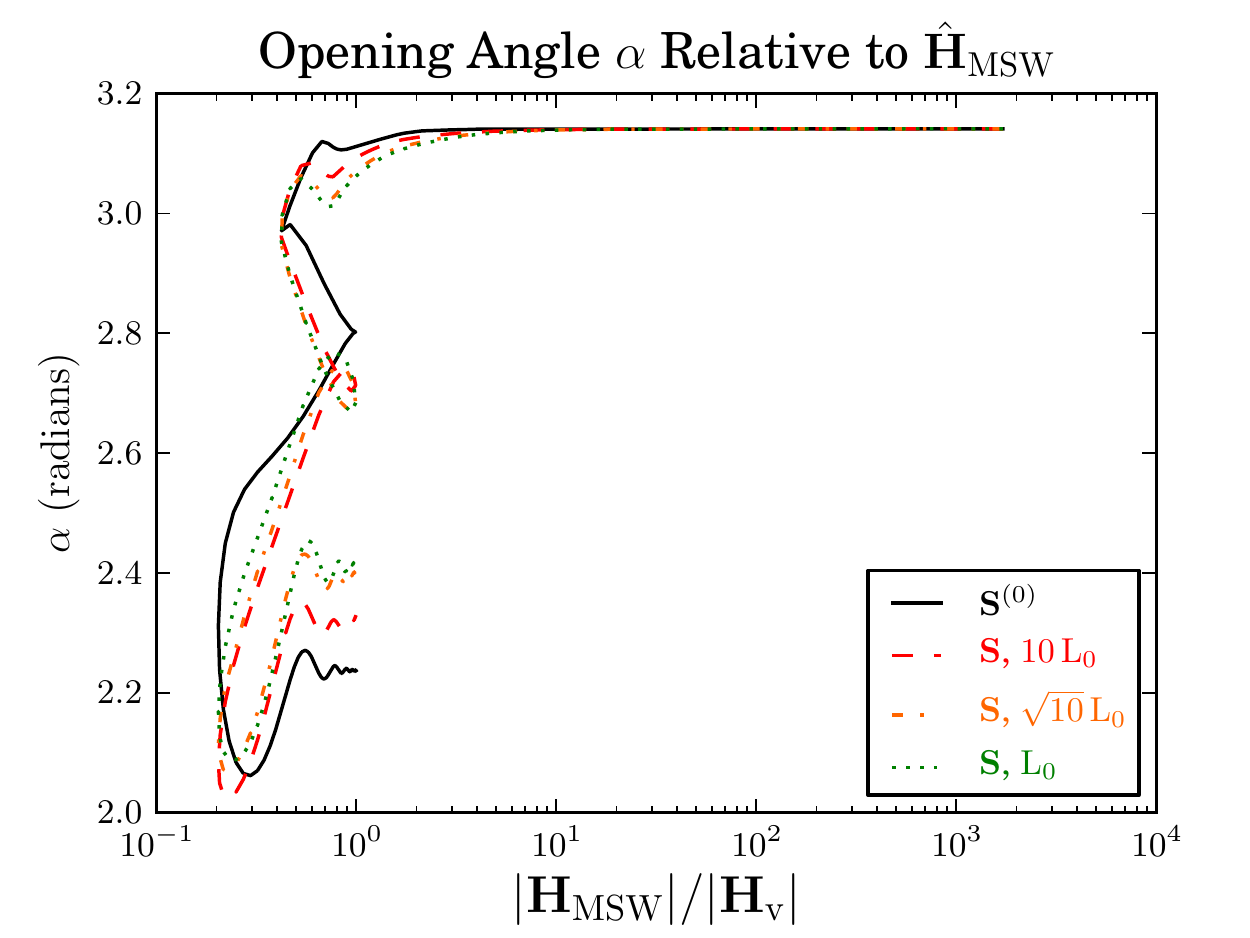}
\caption{High luminosity evolution ($L_0 = 10^{53}\,\rm erg\,\rm s^{-1}$):  The opening angle $\alpha$ between the collective NFIS $\mathbf S^{(0)}$ and $\mathbf H_{\rm MSW}$, plotted as a function of $\left|\mathbf H_{\rm MSW}\right|/\left|{\mathbf H}_{\rm v}\right|$ as the system moves through resonance.  The idealized NFIS (solid line) shows the evolution of $\mathbf S^{(0)}$ in the ideal, strong neutrino self-coupling case.  The dashed line, dot-dashed line, and dotted line show the evolution of $\mathbf S$ as calculated for neutrino luminosities $10L_0,\ \sqrt{10}L_0,\ \rm and\ L_0$ respectively.}
\label{fig:ALGNHIGH}
\end{figure}

Furthermore, equations~\ref{S0} and~\ref{NBEEOM} also imply that a sufficiently large neutrino-neutrino scattering potential will cause neutrinos with oscillation frequencies roughly equal to or less than $\langle\omega\rangle$ to follow the evolution of $\mathbf S^{(0)}$ as this vector moves through resonance.  To illustrate this, we choose to define the angle $\alpha$ as the angle between $\mathbf H_{\rm MSW}$ and either $\mathbf S^{(0)}$ or $\mathbf S$.  The heavy mass eigenstate survival probability $P_{\rm H}$ of the collective ensemble of neutrinos that follow the evolution of $\mathbf S^{(0)}$ is related to $\alpha$ by 
\begin{equation}
P_{\rm H} = 1 - \frac{1}{2}\left(1+\cos{\alpha}\right) .  
\label{PH}
\end{equation}
Figure~\ref{fig:ALGNHIGH} shows the evolution of $\alpha$ for the simulations with relatively high luminosities.  From the figure it can be seen that the evolution of $\mathbf S$ for these luminosities is qualitatively similar to that of $\mathbf S^{(0)}$.  

Interestingly, the final alignment angle, $\alpha$, for the collective neutrino isospin vectors is slightly larger than it is for $\mathbf S^{(0)}$.  This means that the collective NFIS's are more closely aligned with $-\hat{\mathbf H}_{\rm v}$ than $\mathbf S^{(0)}$ is.  The highest luminosity simulation, with $L_{\nu} = 10^{54}\,\rm erg\,\rm s^{-1}$,  has the collective NFIS that is most closely aligned with $\mathbf S^{(0)}$, and the reason for this can be found in Eq.~\ref{NBEEOM}.  In the limit of $\mu\left( r\right) \gg \vert\mathbf H_{\rm e}\vert \gg \omega$ the individual $\mathbf s_{\omega}$ will orbit exclusively around $\mathbf S^{(0)}$ and follow it through resonance.  However, it can be seen from Eqs.~\ref{cos2thm} and~\ref{Homega} that only neutrinos with $\omega = \langle \omega \rangle$ go through resonance at the exact position where $\cos{2\theta_{\rm m}} = 0$.  Neutrinos following the evolution of $\mathbf S^{(0)}$ will still experience some fraction of the neutrino self-coupling potential, although at resonance $\vert\mathbf H_{\rm e}\vert \gg B$ for neutrinos that track $\mathbf S^{(0)}$.  This results in a small increase in ${\cal H}_{\rm res}$, which slightly lowers the overall hopping probability and slightly increases $\alpha$.

\begin{figure}[htp]
\centering
\includegraphics[scale=0.7]{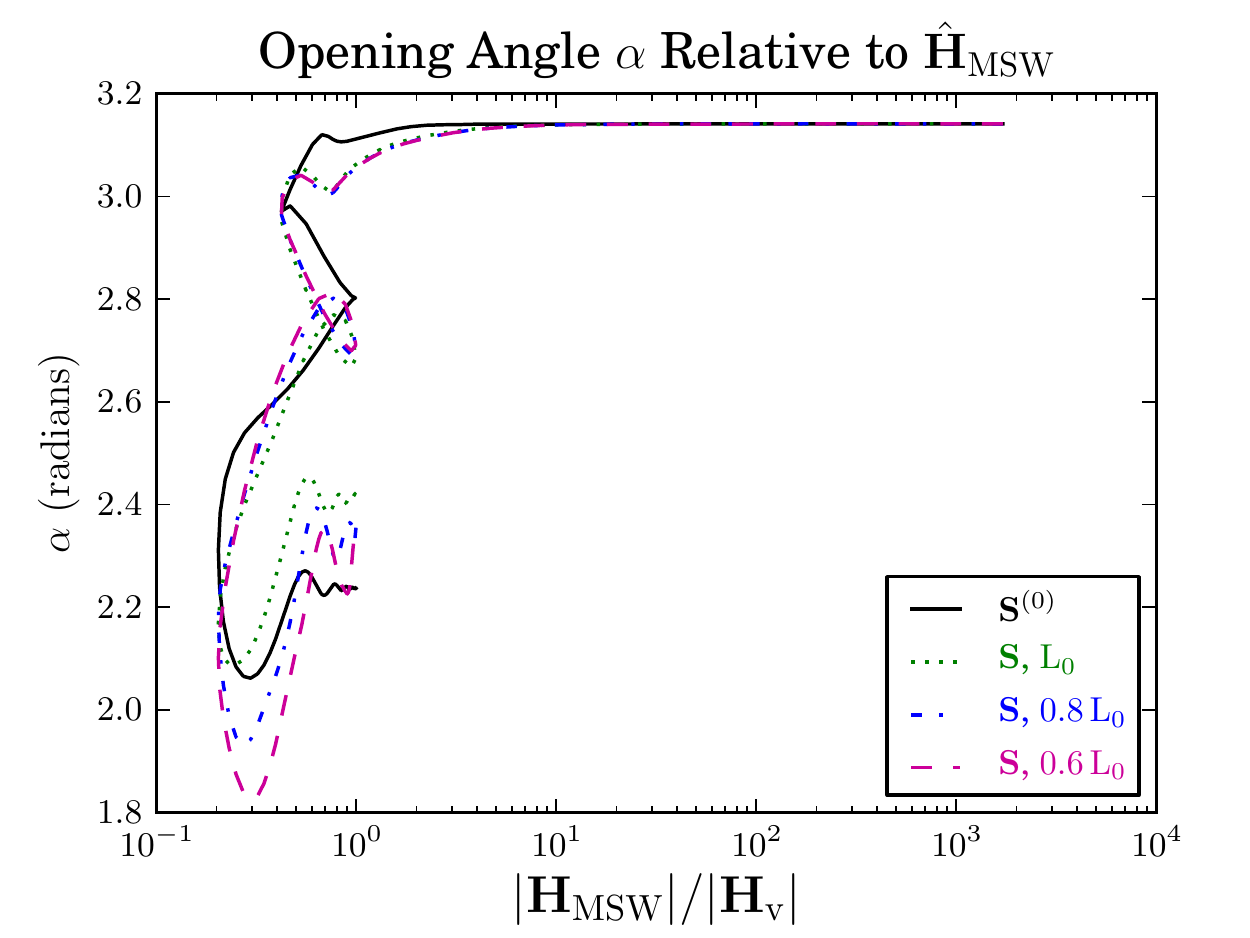}
\caption{Moderate luminosity evolution ($L_0 = 10^{53}\,\rm erg\,\rm s^{-1}$):  The opening angle $\alpha$ between the collective NFIS $\mathbf S^{(0)}$ and $\mathbf H_{\rm MSW}$, plotted as a function of $\left|\mathbf H_{\rm MSW}\right|/\left|{\mathbf H}_{\rm v}\right|$ as the system moves through resonance.  The idealized NFIS (solid line) shows the evolution of $\mathbf S^{(0)}$ in the ideal, strong neutrino self-coupling case.  The dashed line, dot-dashed line, and dotted line show the evolution of $\mathbf S$ as calculated for neutrino luminosities $L_0,\ 0.8\,L_0,\ \rm and\ 0.6\,L_0$ respectively.}
\label{fig:ALGNMID}
\end{figure}

When the neutrino luminosity is moderately lower, the same basic phenomenology is observed.  Figure~\ref{fig:ALGNMID} shows the evolution of $\alpha$ for the simulations with moderate luminosities, $0.6-1.0\times 10^{53}\,\rm erg\,\rm s^{-1}$.  The collective NFIS $\mathbf S$ for these simulations still tracks roughly the evolution of $\mathbf S^{(0)}$, although deviations become more pronounced as the neutrino luminosity decreases.  Counter-intuitively, the final alignment of the lower luminosity $\mathbf S$'s is closer to that of $\mathbf S^{(0)}$ than in the calculations with $L_\nu = 10^{53}\,\rm erg\,\rm s^{-1}$.  This effect originates in the contribution of the bump-affected neutrinos to the integrals in Eqs.~\ref{Sdef} and~\ref{dSdef}.  From Figure~\ref{fig:AllLum} one can see that the population of bump-affected neutrinos has grown appreciably in this luminosity range, comprising $7-10\,\%$ of all neutrinos.  These bump affected neutrinos are not connected in a coherent fashion to the flavor evolution of $\mathbf S$, but they are predominantly aligned with the $+\hat{\mathbf H}_{\rm v}$ axis (they are predominantly $\nu_2$).  This means that they will tend to drag the alignment of $\mathbf S$ closer to the $+\hat{\mathbf H}_{\rm v}$ axis, which systematically moves the final value of $\alpha$ lower.

\begin{figure}[htp]
\centering
\includegraphics[scale=0.7]{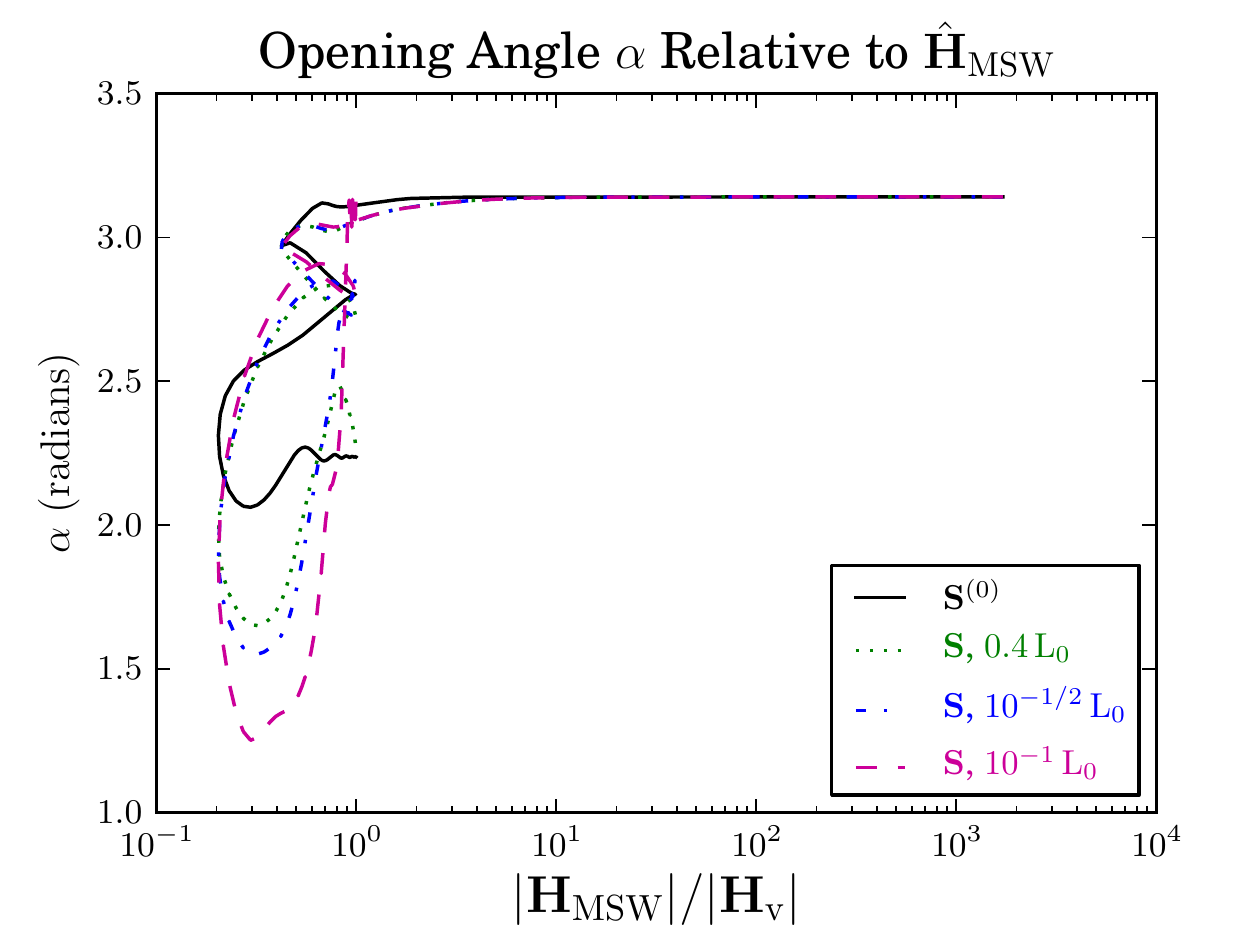}
\caption{Low luminosity evolution ($L_0 = 10^{53}\,\rm erg\,\rm s^{-1}$):  The opening angle $\alpha$ between the collective NFIS $\mathbf S^{(0)}$ and $\mathbf H_{\rm MSW}$, plotted as a function of $\left|\mathbf H_{\rm MSW}\right|/\left|{\mathbf H}_{\rm v}\right|$ as the system moves through resonance.  The idealized NFIS (solid line) shows the evolution of $\mathbf S^{(0)}$ in the ideal, strong neutrino self-coupling case.  The dashed line, dot-dashed line, and dotted line show the evolution of $\mathbf S$ as calculated for neutrino luminosities $0.4\,L_0,\ 10^{-1/2}\,L_0,\ \rm and\ 10^{-1}\,L_0$ respectively.}
\label{fig:ALGNLOW}
\end{figure}

For luminosities below $L_\nu = 6.0\times 10^{52}\,\rm erg\,\rm s^{-1}$, the magnitude of the neutrino self-coupling potential drops below $\langle \omega \rangle = 1.4\times10^{-16}\,\rm MeV$ prior to reaching the resonance region of $\mathbf S^{(0)}$.  This means that many neutrino states will undergo an MSW-like flavor transformation prior to reaching this region of the supernova envelope.  As a result, these neutrinos, including those with $\omega=\langle\omega\rangle$, will not participate in the collective flavor oscillations we have described.  In turn, this means that the approximation $g\left(\omega\right)\approx\delta\left(\omega - \langle\omega\rangle\right)$ is not justified in this case.  Ultimately this approximation breaks down because the evolution of the the neutrino state with $\omega = \langle\omega\rangle$ is not coherently related to the evolution of other neutrino flavor states.  

The progressive breakdown of this approximation with decreasing neutrino luminosity can be seen in Figure~\ref{fig:ALGNLOW}.  The motion of the vector $\mathbf S$ for each calculation deviates widely from the motion of $\mathbf S^{(0)}$.  Furthermore, the final alignment angle $\alpha$ for each $\mathbf S$ is no longer related to the actual hopping probability for neutrinos in those calculations.  The hopping probability inferred from Figure~\ref{fig:ALGNLOW} and Equation~\ref{PH} differs dramatically from the actual hopping probability observed in the calculations shown in Figures~\ref{fig:AllLum} and~\ref{fig:AllPH} for low $L_{\nu}$.  The discrepancies are $\Delta P_{\rm hop} = 0.16,\ 0.35,\ 0.48$ for the calculations with $L_\nu = 4.0,\ \sqrt{10},\ 1.0\times 10^{52}\,\rm erg\,\rm s^{-1}$ respectively.

\subsection{Spectral Swap Formation}
If the luminosity is large enough, it can be seen that after the neutrinos in these calculations have passed the resonance region, the neutrino self-coupling potential becomes the dominant term in the neutrino forward scattering potential.  Neutrinos which have $\omega < \mu\left( r\right)$ fall into a form of collective flavor oscillations known as the Regular Precession mode.  The Regular Precession mode is typified by the collective ensemble of neutrinos rotating with a common frequency, $\omega_{\rm pr}$ about the axis of the vacuum mass basis,
\begin{equation}
\frac{d}{dr}\mathbf s_\omega = \omega_{\rm pr}\left(\mathbf s_\omega \times \mathbf H_{\rm v}\right).
\label{RPEOM}
\end{equation}
This collective oscillation has the feature that it conserves an effective lepton number (or \lq\lq energy\rq\rq ) of the ensemble of neutrinos.  While this lepton number has a more complicated general expression, in the particular case of the neutronization neutrino burst where the initial flux of neutrinos is nearly pure $\nu_{\rm e}$, the conserved lepton number is simply $\propto n_{\nu}P_{\rm H}$ for neutrino mixing at the atmospheric mass scale.  

In the initial stages of neutrino flavor transformation, this lepton number is not conserved.  However, thereafter the Regular Precession mode fixes the total number of neutrinos in mass state 3.  This gives the criterion for the precession frequency, $\omega_{\rm pr}$, for the system,
\begin{equation}
\int^{\infty}_{\omega_{\rm pr}}P_H\left(\omega\right) g\left(\omega\right)d\omega \\
= \int_{0}^{\infty}  g\left(\omega\right)\{P_H\left(\omega\right)-\\
\left[P_2\left(\omega\right) + P_1\left(\omega\right)\right]\}d\omega ,
\label{wprdef}
\end{equation}
where $P_1\left(\omega\right),\ P_2\left(\omega\right)$ are the probabilities of a neutrino with oscillation frequency $\omega$ to be in the instantaneous mass eigenstate 1, or 2 respectively.

It is this precession frequency that sets the energy of the spectral swap (in this case between mass state 3 and mass state 2).  As the magnitude of the self coupling drops, neutrinos with oscillation frequencies in the range $\mu\left( r\right) > \omega > \omega_{\rm pr}$ will participate in the the Regular Precession mode and will align with mass state 3, while neutrinos with $\omega < \omega_{\rm pr}$ will be aligned with mass state 2.  The final results of this process can be seen in Figure~\ref{fig:AllLum}.  For all but the least luminous calculation a spectral swap forms, with $E_{\rm swap} = \delta m^2/2\omega_{\rm pr}$, close to $E_\nu \sim 15\, \rm MeV$.  The precise location of $E_{\rm swap}$ depends on the details of flavor transformation due to the motion of $\mathbf S^{(0)}$ and the bump affected neutrinos.  Broadly speaking, a smaller $P_{\rm H}$ found from Eq.~\ref{PH} will lower the swap energy of the final neutrino energy spectra by reducing the value of the integral on the right side of Eq.~\ref{wprdef}.  However, a larger population of bump affected neutrinos will move the swap energy to higher values (smaller $\omega_{\rm pr}$) by reducing $P_{\rm H}\left(\omega\right)g\left(\omega\right)$ for large $\omega$.

It is important to note that the spectral swap between mass states 3 and 2 can sometimes form even when the coherent flavor evolution of neutrinos has broken down deeper in the envelope.  As discussed in the previous section, for the calculations with $L_\nu < 6.0 \times 10^{52}\,\rm erg\,\rm s^{-1}$ collective neutrino flavor transformation breaks down in the resonance region because $\mu\left( r\right) < \langle\omega\rangle$.  However, from Figure~\ref{fig:AllLum} it can be seen that a mass state 3/2 swap still forms successfully for $L_\nu = 4.0 \times 10^{52}\,\rm erg\,\rm s^{-1}$ and $L_\nu = \sqrt{10} \times 10^{52}\,\rm erg\,\rm s^{-1}$.  Swaps between mass state 3 and 2 form for these two models where the luminosity is low because the neutrino self-coupling is still large compared to $\omega$ for high energy neutrinos, specifically $\mu\left( r\right) > \omega_{\rm pr}$ after the resonance region.  This allows the high energy neutrinos to briefly form a Regular Precession mode before $\mu\left( r\right)$ decreases further with radius and flavor transformation in the $\delta m^2_{\rm atm}$ mixing sector stops.  

The swaps are incomplete for the calculations with $L_\nu = 4.0 \times 10^{52}\,\rm erg\,\rm s^{-1}$ and $L_\nu = \sqrt{10} \times 10^{52}\,\rm erg\,\rm s^{-1}$.  The calculations for these cases show small populations of $\nu_2$ neutrinos below $E_{\rm swap}$ which are not bump affected.  They also show small populations of neutrinos in $\nu_3$ with energies above $E_{\rm swap}$.  This phenomenon arises because of the extremely short lifetime of the collective precession at these luminosities.  We define $r_{\rm stop}$ to be the distance between the end of the resonance region and the point at which $\mu\left( r\right) = \omega_{\rm pr}$.  Comparing that to the oscillation length of neutrinos in the regular precession mode, $l_{\rm osc} = 2\pi /\omega_{\rm pr}$, we find that for these two cases $r_{\rm stop} \sim l_{\rm osc}$.  Clearly, a spectral swap cannot fully form if the Regular Precession mode ceases before it can complete several full oscillations.  It is interesting, however, that the swaps in these two calculations are as pronounced as they are given the rapid truncation of the collective neutrino oscillations.

For the lowest neutrino luminosity, $L_\nu = 1.0 \times 10^{52}\,\rm erg\,\rm s^{-1}$, there is no spectral swap observed in Figure~\ref{fig:AllLum} between mass states 3 and 2.  In this case, $\mu\left( r\right) < \omega_{\rm pr}$ even before the system finishes MSW-like flavor transformation.  No collective oscillation can proceed in the $\delta m^2_{\rm atm}$ mixing sector for this case.

While we have focused entirely on the $\delta m^2_{\rm atm}$ mixing sector in the sections above, it should be pointed out that the $\delta m^2_{\odot}$ mixing sector is completely indifferent to the range of luminosities that we have explored.  The neutrino flavor mixing energy scale for the $\delta m^2_{\odot}$ mass state splitting is $\sim 30$ times smaller than that of the atmospheric mass state splitting.  The model of neutrino flavor transformation outlined above is quite robust for the solar mixing sector, with $\mu\left( r\right) > \langle\omega\rangle_{\odot}$ and $\mu\left( r\right) > \left(\omega_{\rm pr}\right)_{\odot}$ for all of the neutrino luminosities that we consider.  The spectral swap between mass states 2 and 1 is created by the Regular Precession mode in this mixing sector.  While ${E_{\rm swap}}_\odot$ varies greatly for the different calculations in Figure~\ref{fig:AllLum}, this swap energy is only changed by variations in flavor transformation in the $\delta m^2_{\rm atm}$ sector.  The ratio of $\nu_2 / \nu_1$ neutrinos is identical for all of the calculations shown in Figure~\ref{fig:AllLum}.  A curious consequence of this is that the spectral swap energies move closer and closer together as the luminosity of the neutronization burst decreases, until ultimately the swap between mass state 3 and 2 disappears altogether.  This behavior is evident in Figure~\ref{fig:AllLum}.

\section{Discussion and Conclusions}
We have shown that the bump in the electron number density profile 
at the base of the hydrogen envelope in O-Ne-Mg core-collapse 
supernovae causes an interesting interplay 
between neutrino-electron and neutrino-neutrino 
forward scattering effects in the flavor evolution of low-energy $\nu_e$ 
during the neutronization burst epoch.
The bump allows a significant fraction of the low-energy $\nu_e$ 
to remain in the electron flavor state, i.e., enhancing their survival probability.  It does this by rendering their flavor evolution nonadiabatic.
Additionally, we have found that increasing the luminosity $L_\nu$ of the neutronization burst
shifts the bump-affected $\nu_e$ to lower energy with consequently reduced
survival probability. Finally, we have found that the flavor states of the bump affected low energy neutrinos impact the spectral swap forming behavior of the collective oscillations later on.  This opens up the possiblity that the presence of bump affected low-energy $\nu_e$ of $\sim 3$--5 MeVcan be detected through the spacing of the spectral swaps.  However, this may not be necessary as some neutronization burst signals may produce bump affected neutrinos at energies that are accessible by Earth based detectors.  This may also prove to be an interesting secondary probe of the burst luminosity $L_\nu$ of an O-Ne-Mg core-collapse supernova.

While the zeroth-order mean field proves to be rather useful
in understanding the flavor evolution of the low-energy $\nu_e$, it is
clearly inadequate in providing a quantitative description of the flavor
evolution if the $\nu_e$ burst luminosities are on the low side of the expected emission (see Figure~\ref{fig:ALGNLOW}).  In particular, it cannot
provide a good estimate for the
energy at which the spectral swap occurs in the low luminosity limit. We note that 
the spectral swap occurs rather robustly at a fixed energy once
$L_\nu$ exceeds $\sim \sqrt{10}\times 10^{52}$~erg/s (see Figure~\ref{fig:AllLum}).





\section{Acknowledgments}
This work was supported in part by NSF grant PHY-06-53626 at UCSD, DOE grant DE-FG02-87ER40328 at the UMN, and by the DOE Office of Nuclear Physics, the LDRD Program and Open Supercomputing at LANL, and an Institute of Geophysics and Planetary Physics/LANL minigrant.  We would like to thank the topical collaboration for neutrino and nucleosynthesis in hot and dense matter at LANL and the New Mexico Consortium for providing a stimulating platform to carry out this work.

\bibliography{allref}
\end{document}